\documentclass[a4paper,11pt]{article}
\usepackage{jinstpub} 
\usepackage{graphicx}
\usepackage{sidecap}
\usepackage{wrapfig}
\usepackage{float}
\usepackage{supertabular}
\usepackage{array}
\usepackage{threeparttable}
\usepackage{booktabs}
\usepackage{setspace}
\usepackage{url}
\usepackage{amsmath}
\usepackage{amsfonts}
\usepackage[version=3]{mhchem}  
\usepackage{indentfirst}
\usepackage{subcaption}
\usepackage{caption}
\usepackage{pdflscape}
\usepackage{lipsum}
\usepackage{blindtext}
\usepackage[nottoc]{tocbibind}
\usepackage{pgfplots}
\usepackage{pgfplotstable}
\pgfplotsset{compat=1.7}
\usepackage{tikz}
\usepackage{xcolor}
\usepackage{etoolbox}
\usepackage{flafter}
\usepackage{placeins}
\usepackage{booktabs} 
\usepackage{capt-of} 
\usepackage{graphicx} 
\usepackage{cleveref}

\newcommand{\orcid}[1]{\href{https://orcid.org/#1}{\includegraphics[height=1em]{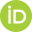}}}

\title{\centering \Huge\bf Jet Image Tagging Using Deep Learning: An Ensemble Model}

\author[a,1]{Juvenal Bassa\orcid{0009-0006-1061-0680} \note{Corresponding author}}
\author[b]{, Vidya Manian\orcid{0000-0003-3834-8857}}
\author[a]{, Sudhir Malik\orcid{0000-0002-6356-2655}}
\author[a]{, Arghya Chattopadhyay\orcid{0000-0002-6427-5076}}

\affiliation[a]{Department of Physics, University of Puerto Rico Mayaguez}
\affiliation[b]{Department of Electrical and Computer Engineering, University of Puerto Rico Mayaguez,\\ 
PR 00681, USA}

\emailAdd{juvenal.bassa@upr.edu}
\emailAdd{vidya.manian@upr.edu}
\emailAdd{sudhir.malik@upr.edu}
\emailAdd{arghya.chattopadhyay@upr.edu}

\abstract{Jet classification in high-energy particle physics is important for understanding fundamental interactions and probing phenomena beyond the Standard Model. Jets originate from the fragmentation and hadronization of quarks and gluons, and pose a challenge for identification due to their complex, multidimensional structure. Traditional classification methods often fall short in capturing these intricacies, necessitating advanced machine learning approaches. In this paper, we employ two neural networks simultaneously as an ensemble to tag various jet types. We convert the jet data to two-dimensional histograms instead of representing them as points in a higher-dimensional space. Specifically, this ensemble approach, hereafter referred to as Ensemble Model, is used to tag jets into classes from the JetNet dataset, corresponding to: Top Quarks, Light Quarks (up or down), and \textit{W} and \textit{Z} bosons. For the jet classes mentioned above, we show that the Ensemble Model can be used for both binary and multi-categorical classification. This ensemble approach learns jet features by leveraging the strengths of each constituent network achieving superior performance compared to either individual network.}


\begin{document}

\maketitle

\section{Introduction} \label{sec:S1}
\noindent The Large Hadron Collider (LHC), located at CERN (European Organization for Nuclear Research), is designed to explore the fundamental constituents of matter and their nature. It collides protons at a total center-of-mass energy of 13 TeV, allowing scientists to precisely probe the Standard Model and search for physics beyond it. There are several detectors located at the LHC, and among them, CMS and ATLAS are the two general-purpose detectors. These are designed to detect and measure the energy and momentum of a wide range of particles produced in these high-energy collisions \cite{cms2008cms, abbrescia2008cms}. In understanding the fundamental interactions and the properties of particles produced in high-energy collisions, jets - narrow cones of hadrons and other particles - play a critical role. Jets are formed when high-energy quarks or gluons produced during particle collisions undergo fragmentation and hadronization, resulting in cascades of stable particles, such as pions and kaons \cite{kogler2019jet, dreyer2018lund}. Jets play a crucial role in probing the fundamental laws of nature, particularly Quantum Chromodynamics (QCD), which describes the interactions between quarks and gluons. They are indispensable in exploring the Standard Model and beyond, offering insights into phenomena like Higgs boson production, top quark studies, and searches for new physics. Understanding jets is also vital for interpreting events that involve large missing transverse energy, which may signal new particles \cite{collaboration2012observation}. To detect jets, the LHC detector employs sophisticated tracking systems, calorimeters, and muon chambers to capture the properties of the constituent particles. Particles comprising jets leave signals in components such as the tracker and the electromagnetic and hadron calorimeters. These components work in tandem to reconstruct particles belonging to jets by clustering energy deposits from charged and neutral particles. These signals are combined using jet algorithms to form what is called a \emph{reconstructed jet}.  \\

\noindent As is traditional in collider physics, a three-dimensional Cartesian coordinate system is used, with the $z$-axis defined as the direction of the beam axis. The transverse momenta of the particles along the $(x,y)$ plane are denoted by $p_T$. As another common practice, rather than using particle momenta in Cartesian coordinates as $(p_x,p_y,p_z)$, the kinematic information for each resulting particle is stored in terms of the $(p_T,\eta, \phi)$ coordinate, where $\phi$ is calculated as the angle of the particle trajectory with respect to the $x$ axis and $\eta=-\log(\tan(\theta/2))$ is the pseudo-rapidity with $\theta$ being the polar angle. Since each jet is a collection of different particles, a single jet can either be represented as a cloud of points in the kinematic space of its constituents, or as a single vector in a higher-dimensional feature space by concatenating the particle-level information.\\

\noindent The distinction between different types of jets (e.g., gluon and quark jets), resulting from high-energy processes, is a challenging task due to the subtle structural features involved in the jet data. Jet tagging, the process of classifying and identifying jets correctly, is vital to the success of the LHC physics program. It involves classifying jets based on their originating particle, such as distinguishing between quark and gluon jets or identifying those arising from heavier particles like the Higgs boson or the top quark. Accurate jet tagging is essential for isolating signals of interest in collider data, reducing background, and enhancing the discovery potential for new physics \cite{larkoski2020jet}. Recent advances in Machine Learning (ML) for jet tagging are presented in \cite{schwartz2021modern}. Deep Learning (DL) models, such as ParticleNet \cite{Qu_2020}, the Particle Transformer \cite{qu2022particle}, and Lorentz Equivariant Network have demonstrated exceptional accuracy \cite{gong2022efficient}. These ML-based methods surpass traditional approaches by learning complex, non-linear features directly from data. They leverage rich representations of jets, including particle-level features and interactions. However, they require a large amount of training data and high computational power. For example, Particle Transformer \cite{qu2022particle} and ParticleNet \cite{qu2020jet} are leading architectures that operate on particle-level inputs and perform well on benchmark tasks. Trained on $100$ million jets from the JetClass dataset \cite{Kansal_MPGAN_2021}, ParticleNet and Particle Transformer achieve accuracies of $84.4\%$ and $86.1\%$, respectively, when classifying all nine jet classes available in that dataset. A key motivation behind our Ensemble Model is to achieve comparable performance using significantly less training data. In contrast to JetClass, we train on the JetNet dataset \cite{Kansal_MPGAN_2021, kansal2021particle}, which contains fewer jet features and only five jet classes. Despite this, our model reaches around $75\%$ classification accuracy using fewer than $1$ million jets, demonstrating that the ensemble approach can efficiently learn discriminative features even from limited and less rich data. However, we note that a direct numerical comparison between models trained on JetNet and JetClass is not appropriate due to differences in dataset richness.\\

\newpage
\noindent Given the complexity of the jet data, jets can be visualized in multiple ways, including but not limited to viewing jets as images \cite{atlas2017quark, cogan2015jet, de2016jet,deOliveira:2015xxd}. In these cases, the particle constituents of each jet are projected onto a discretized $\eta-\phi$ plane, by taking the intensity of the pixels in this grid to be a monotonically increasing function of $p_T$. These images can be thought of as the projection of particle showers in a $2D$ plane. One of the advantage of viewing jets as images is that the image itself can be viewed as some form of geometric data describing the \emph{visual structure} of the jet. As discussed before, the origin of a jet lies in the intrinsic properties of some particles, therefore the shape or geometry of these showers encode some of the intrinsic properties of the source particle itself. One of the motivations to start with jet images as our initial point is exactly to exploit the hidden geometric symmetries of these showers for aiding classification. While extensive literature exists on DL for jet tagging, particularly for jet image representation, the application of convolutional neural network (CNN)-based models for point cloud jet image classification remains an area of active research \cite{cogan2015jet, guest2018deep}. As we will motivate in the following Section, CNNs are excellent tools to exploit symmetries hidden in the data or image. Therefore to exploit the symmetries or sub-structures of jet images as well as 
to minimize the data dependency and faster training, all the while keeping high value of accuracy we are taking two main steps in this paper:

\begin{itemize}
    \item Following \cite{kansal2021particle}, jets are converted into images by visualizing the distribution of relative transverse momenta, $p_T^{rel}(=p_T^{\text{particle}}/p_T^{\text{jet}})$ of constituent particles as the intensity of pixels in the $(\eta^{rel}, \phi^{rel})$ plane where $\phi^{\text{rel}}(=\phi^{\text{particle}}-\phi^{\text{jet}}\,(\text{mod }2\pi))$ and $\eta^{\text{rel}}(=\eta^{\text{particle}}-\eta^{\text{jet}})$. This transformation allows one to use DL models originally developed for image classification to be developed or adapted for jet tagging.
    \item Application of two pre-trained networks\footnote{Networks that are already trained for a different but related task.}, with distinct CNN-based network architectures namely, the \emph{ResNet50} \cite{he2016deep} and the \emph{InceptionV3} model \cite{szegedy2016rethinking}, simultaneously in an ensemble to effectively combine and leverage the unique strengths of each model  \cite{rokach2010ensemble}. This combination is referred to as the Ensemble Model (EM) for the rest of the paper.
\end{itemize}

\noindent To support training, validation and testing of the EM architecture, JetNet dataset \cite{Kansal_MPGAN_2021, Kansal_JetNet_2023} is used, a public benchmark that provides point cloud representations of jets from simulated proton-proton collisions. This dataset, suitable for both binary and multi-class classification tasks, includes labeled jets from key particle origins such as gluons, light quarks(up and down), top quark and heavy bosons like $W$ and $Z$ boson. The full pipeline of codes from generating images to implementing the training of EM is available at \href{https://github.com/Basjuven/Point-Cloud-jet-images-tagging-using-DL/blob/main/JetNet_GPU_image_gen.ipynb}{this} GitHub repository\cite{Juvbassa}. \\

\noindent This paper is organized as follows. Section \ref{sec:S2} introduces the relevant features of the JetNet dataset and the pre-processings relevant for this work. In Section \ref{sec:S3}, methodology and architecture of EM is described along with its training protocol on the JetNet dataset. Section \ref{sec:S4} presents the main results of the present study. Finally Section \ref{sec:S5} consists the conclusion and discussions about this study. In addition, appendix \ref{append:S1} is included for summarizing the architecture of each of the two models used as part of EM.

\section{Dataset and jet image preparation} \label{sec:S2}

\begin{figure} [H] 
    \centering
    \begin{subfigure}{0.32\textwidth}
        \centering
        \includegraphics[width=\textwidth]{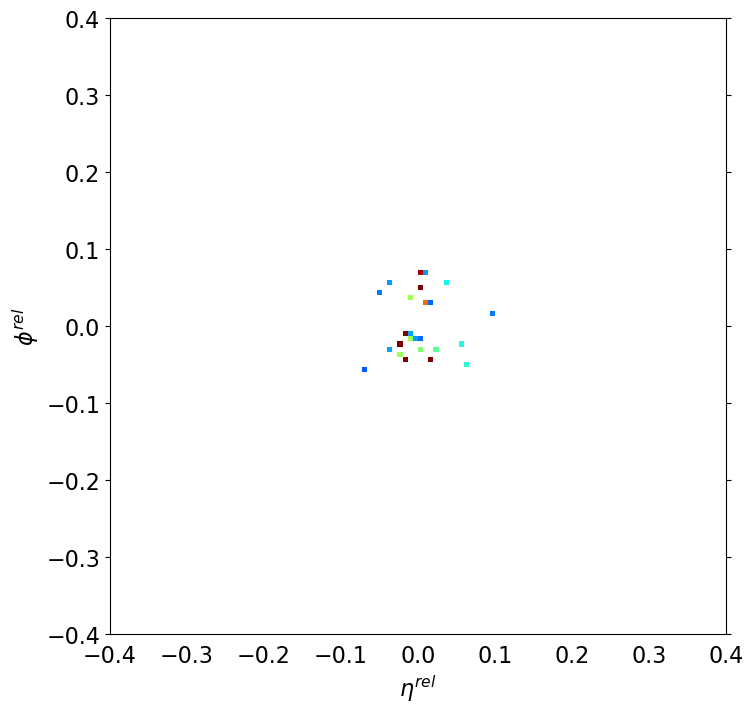} 
        \caption{$g$-jet: derived from gluon.} 
        \label{fig:image1}
    \end{subfigure}
    \hfill
    \begin{subfigure}{0.32\textwidth}
        \centering
        \includegraphics[width=\textwidth]{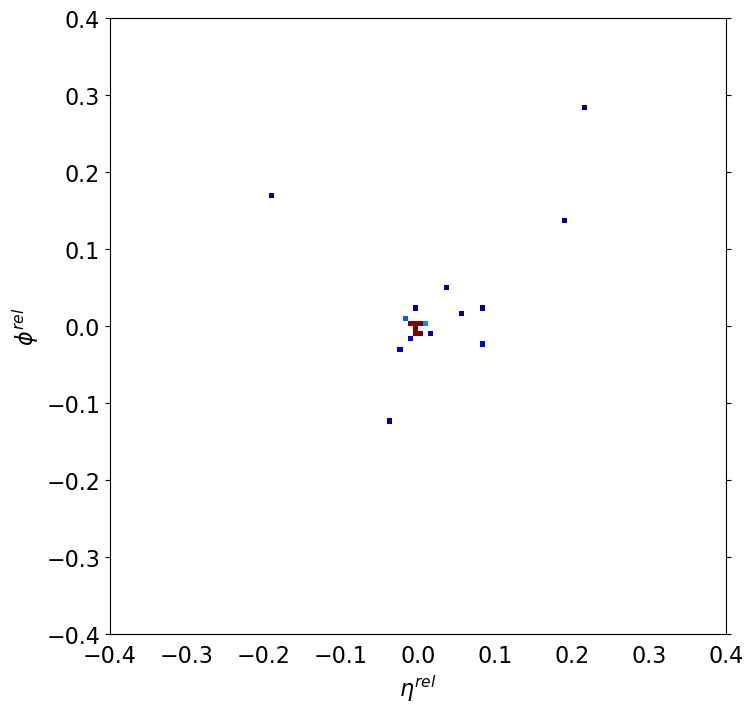} 
        \caption{$q$-jet: derived from quark}
        \label{fig:image2}
    \end{subfigure}
    \hfill
    \begin{subfigure}{0.32\textwidth}
        \centering
        \includegraphics[width=\textwidth]{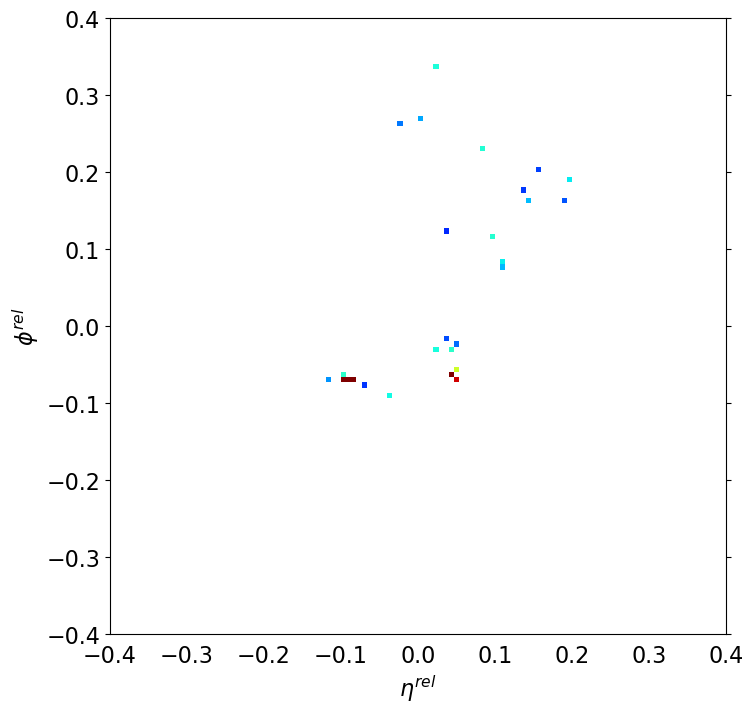} 
        \caption{$t$-jet: derived from top quark}
        \label{fig:image3}
    \end{subfigure}
    
     \vspace{0.5cm}
    \begin{subfigure}{0.32\textwidth}
        \centering
        \includegraphics[width=\textwidth]{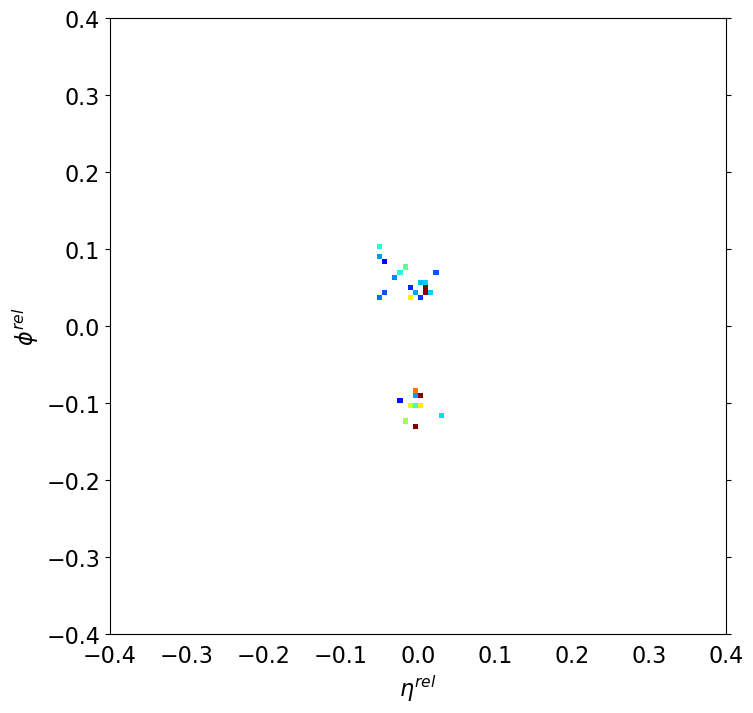} 
        \caption{$W$-jet: derived from $W$ boson}
        \label{fig:image4}
    \end{subfigure}
    \begin{subfigure}{0.32\textwidth}
        \centering
        \includegraphics[width=\textwidth]{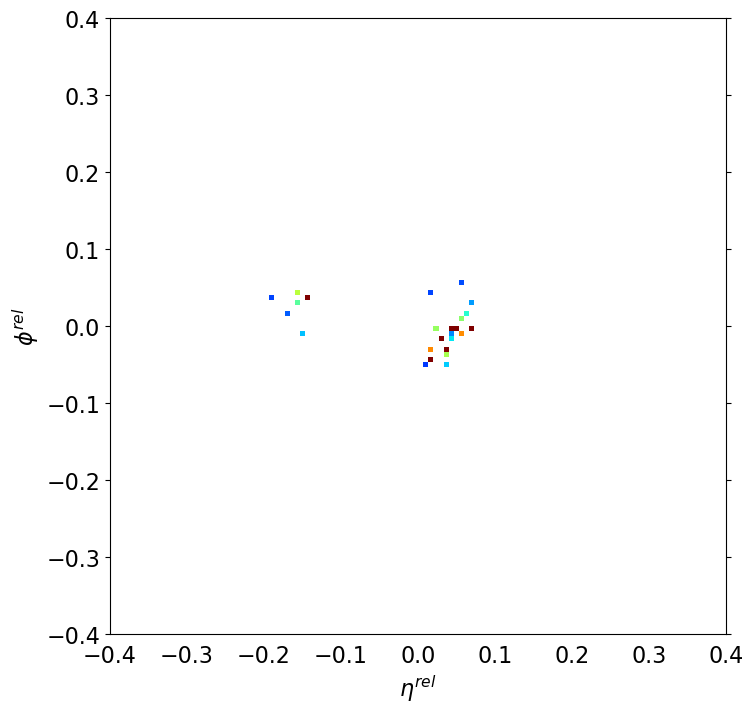} 
        \caption{$Z$-jet: derived from $Z$ boson}
        \label{fig:image5}
    \end{subfigure}
    
    \caption{Jets representation as two-dimensional images, constructed from point cloud JetNet data using the anti-$k_t$ jet clustering algorithm.}
    \label{fig:jets_images}
\end{figure}

\noindent The JetNet dataset, introduced by \textit{Kansal et al.} in 2021 \cite{Kansal_MPGAN_2021, kansal2021particle}, was developed to support machine learning research in high-energy physics, specifically jet tagging from simulated collision data. It consist of point cloud representations of jets derived from proton-proton ($pp$) collisions at a center of mass energy of $13~\mathrm{TeV}$, constructed using the anti-$k_t$ jet clustering algorithm with a radius parameter of $R=0.4$ \cite{Cacciari:2008gp}. The information is stored in a hierarchical file structure for convenience. The global kinematic information about each jet\footnote{ For each jet, JeTNet consists of $5$ global features namely [$type$, $p_T$, $\eta$, $mass$, $num\_particles$].} includes the type (or the origin) of the jet and the number of particles that constitute this jet. At the next level the information about all the particles in the jet is represented as a collection of particles with individual $\eta^{\text{rel}}$, $\phi^{\text{rel}}$ and $p_{T}^{\text{rel}}$ values, making the dataset particularly suitable for geometric deep learning and graph-based models. The jets are labeled according to their originating particle type, including gluons ($g$-jet), light quarks ($q$-jet), top quarks ($t$-jet), $W$ bosons ($W$-jet), and $Z$ bosons ($Z$-jet). We kept a maximum of $30$ highest $p_T$ particles per jet from the JetNet dataset\footnote{Apart from the kinetic information, the particle level data from JetNet also have a binary "mask" feature classifying the particle as genuine or zero-padded in case some of the jets have less than $30$ particles.}. In this work, by jet image we effectively mean a binned histogram in the $(\eta^{rel}, \phi^{rel})$ plane (figure \ref{fig:jets_images}). The resulting images encode the spatial energy distribution of the jet constituents, with pixel intensity proportional to the $p_{T}^{\text{rel}}$ of each particle or the total $p_{T}^{\text{rel}}$ deposited within a given bin. At this point, one should note two of the most critical choices that one needs to make before representing jets as images. Both of which primarily rely on the fact that any form of discretizations of data would inevitably result into information loss and one would like to preserve information as much as possible. As our first choice, we would restrict ourselves to the range of data for which $\eta^{\text{rel}} \in [-0.4,0.4]$ and $\phi^{\text{rel}}\in [-0.4,0.4]$. Taking into account the full range of $\eta^{\text{rel}}\in [-1.6,1.0]$ and $\phi^{\text{rel}}\in [-0.5,0.5]$ available in the JetNet dataset would result into most of the images being too sparse to work with. Another crucial choice is the number of bins along the $\eta^{\text{rel}}$ as well as the $\phi^{\text{rel}}$ direction. Too big of a bin width would result into less number of particles being registered in the image and on the other side choosing too small a bin size would again make the image sparse. Keeping in mind this trade-off between bin-size and number of particles showing up in the image in consideration and the network architectures we are going to use, we chose the resolution of the jet images as $299 \times 299$ pixels or in other words we keep $299$ bins for both $\eta^{\text{rel}}$ and $\phi^{\text{rel}}$ directions for the purposes of this work.

\noindent Another advantage of moving towards jet images as the data for classification, hides in the fact that the images itself act as a geometric data for the machines to learn from. For example, gluons radiate into large number of particles before going through hadronization, mostly resulting in a single cone of showering particles which would show up in the $(\eta^{\text{rel}},\phi^{\text{rel}})$ plane as a distribution around a single origin. On the other hand, top quarks might decay
into three lighter quarks through an intermediate $W$ boson, which might then produce their own sub-jets, leading to a complex two- or three- pronged distribution. Therefore, transforming jets as images is a very convenient way to understand the jet substructures from the geometry of the images. \\
\noindent These mappings of the high-dimensional particle-level structure of jets into a fixed-size 2-D grid, invites the use of CNNs to capture local and global substructure patterns relevant for jet classification. JetNet dataset provides a limited number of labeled jet images per particle type, the maximum number varies slightly by particle type: \\

\begin{table}[H]
\centering
\begin{tabular}{l c}
\hline
\textbf{Jet Type} & \textbf{Number of Images} \\
\hline
Gluon         & 177,252 \\
Light Quark   & 170,679 \\
Top Quark     & 177,945 \\
$W$ Boson     & 177,172 \\
$Z$ Boson     & 176,952 \\
\hline
\end{tabular}
\caption{Maximum number of jet images available per particle type in the JetNet dataset \cite{jetnetzenodo}.}
\label{tab:jetnet_counts}
\end{table}

\noindent To ensure class balance and maintain statistical uniformity during training/validation and testing, a consistent dataset was established comprising 170,000 jet images per category. This threshold was selected to match the lowest class availability (see Light Quark in table  \ref{tab:jetnet_counts}) while preserving a sufficient sample size for deep learning applications. This balanced dataset ensures that no class disproportionately influences the learning process, which is especially critical in multi-categorical classification tasks, imbalance could bias performance metrics.
The whole dataset was partitioned using a 80/20$\%$ split for training-validation/testing. To also ensure balanced representation of all jet classes across the subsets, the split was performed in a stratified manner, preserving the class distribution in both the training-validation/testing sets. So, each subset contains an equal number of jet images for each particle type, avoiding class imbalance and ensuring consistent performance evaluation during training and testing phases.

\section{Architecture and Training } \label{sec:S3}
\noindent The task of identifying subtle patterns within high-dimensional dataset is increasingly tackled using Deep convolutional neural networks by the computer vision community, leveraging their ability to hierarchically learn spatial and spectral features \cite{he2016deep}. Due to the complex and heterogeneous nature of jets, no single model architecture is uniformly optimal for all tagging tasks. This is reflected in the wide range of machine learning approaches developed for jet classification, as surveyed in \cite{radovic2018machine}. This limitation is particularly evident in distinguishing jet images originating from heavy particles (e.g., top-quark, $W$/$Z$-boson) from those in QCD backgrounds \cite{larkoski2020jet}, where overlapping substructures and class imbalance hinder generalization. This motivates the use of ensemble methods, such as stacked generalization \cite{wolpert1992stacked}, which can leverage complementary strengths of different models to improve classification performance. In the context of jet classification, ensemble methods can help capture the diverse, intricate patterns present in jet images. \\

\noindent In the case of data with inherent spatial features, such as images or time series, CNNs are the method of choice. Unlike fully connected layers, which use different weights for each connection between neurons in adjacent layers, convolutional layers exploit the spatial symmetry of the input by sharing the same weights (between the input and kernel) across the entire input space. This significantly reduces the number of parameters in the model, making it more efficient by leveraging the symmetries present in the input. In real-world data like color images, one can tweak the kernel to span over multiple \emph{input channels} $C_{in}$ (e.g., RGB image format has $C_{in}=3$) by simply thinking the input as a $3D$ tensor $I\in \mathbb{R}^{H\times W\times C_{in}}$, and a convolutional layer with $C_{out}$ numbers of $3D$ kernels $K\in \mathbb{R}^{k\times k\times C_{in}\times C_{out}}$. Therefore, each output channel $c$ will have the feature map
\begin{equation}
    S^c(i,j)=\sum_{d=1}^{C_{in}}\sum_{m=0}^{k-1}\sum_{n=0}^{k-1}K_d^c(m,n)\cdot I_d(i+m,j+n),
\end{equation}
enabling the network to learn more abstract features as $C_{out}$ grows. Furthermore, to improve learning stability and generalization, most of the time, CNNs incorporate \emph{batch normalization}, which normalizes activations across a batch and \emph{pooling layers} which downsample the feature maps. Using these basic concepts one can design different neural networks with CNN layers in between, depending on the local correlations and translational symmetry of the data. Their efficiency arises from weight sharing and the hierarchical composition of patterns across layers.

\subsection{The Ensemble Model} \label{sec:S3.2}
\begin{figure}[H]
\includegraphics[width=9.5cm]{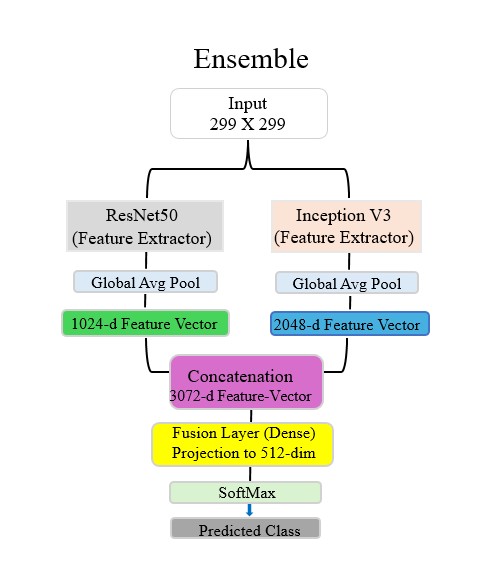}
\centering
\caption{Proposed ensemble model architecture. The model integrates features extracted from \textit{ResNet50} and \textit{InceptionV3}, where a $1024$-dimensional feature vector from \textit{ResNet50} and a $2048$-dimensional feature vector from \textit{InceptionV3} are concatenated, forming a $3072$-dimensional representation. The fusion layer reduces the dimensionality to $512$ before passing it to a softmax classifier.}
\label{fig:ensemble}
\end{figure}

\noindent In our particular implementation, we combine \emph{InceptionV3} and \emph{ResNet50} \cite{szegedy2016rethinking, he2016deep}. While both have convolutional layers in their network structure, they differ significantly in architecture which is elaborated in appendix \ref{append:S1}. \emph{InceptionV3} employs multi-scale convolutional paths with factorized convolutions, while \emph{ResNet50} introduces identity mappings via residual connections. \textit{ResNet50} is well-suited for learning hierarchical and localized patterns, while \textit{InceptionV3} is designed to capture diverse spatial features across multiple scales. By combining their learned feature representations at an intermediate stage, the EM architecture aims to enhance generalization and robustness in both binary and multi-class classification task. In figure~\ref{fig:ensemble}, the architectural design of EM and its fusion strategy to integrate the feature extraction capabilities of two deep learning architectures is represented diagrammatically.
\noindent EM first concatenates $1024$-dimensional feature vector from \textit{ResNet50} (post-global average pooling) and $2048$-dimensional vector from \textit{InceptionV3} (post-average pooling layer). Thereafter a fusion layer reduces dimensionality to $512$ via dense projection, followed by a softmax classifier.

\subsection{Training and validation: K-fold cross-validation} \label{sec:S3.3}
 
\noindent While the weights and biases of the fusion layer in the EM are initialized randomly, both \emph{ResNet50} and \emph{InceptionV3} are initialized with pretrained weights from the ImageNet dataset \cite{deng2009imagenet}. Following the jet image preparation outlined in Section \ref{sec:S2}, the input to the EM consists of the $299 \times 299$ pixel jet images. For training and validation process, the following categorical \textit{cross-entropy loss} function is used along with the \textit{Adam} optimization algorithm \cite{kingma2015adam} for initial learning rate of $l_{r} = 3\times 10^{-4}$:

\begin{equation}
    \mathcal{L} = -\frac{1}{N} \sum_{i=1}^{N} \sum_{c=1}^{C} y_{i,c} \log(p_{i,c}), \label{ec:loss}
\end{equation} 

\noindent where $N$ is the number of samples, $C$ the number of classes, $y_{i,c}$ the true label indicator\footnote{which is 1 if sample $i$ belong to class $c$, 0 otherwise for binary classification.}, and $p_{i,c}$ the predicted probability for class $c$. This loss function penalizes incorrect class predictions with a logarithmic cost and is widely used in multi-class classification tasks due to its probabilistic interpretation and effectiveness in training deep neural networks \cite{bishop2006pattern}. \\

\noindent To ensure robust performance evaluation and reducing variance due to dataset partitioning, $K$-fold cross-validation during model training is applied. $K$-fold cross-validation also assesses the generalization of ML models \cite{arlot2010survey}. To implement cross-validation, the dataset is randomly divided into $K$ equal subsets or folds, with each fold used once as a validation set while the remaining $K-1$ are used for training. This cycle is repeated $K$ times to ensure that all data points contribute to both training and validation \cite{gorriz2024kfold}. The final performance metric is obtained by averaging the results across all iterations, reducing bias and providing a more reliable estimate of the effectiveness of the model \cite{bradshaw2023guide}. All binary and multi-class classification results presented in Section \ref{sec:S4}, where the results are tabulated, follow this cross-validation protocol with the number of folds fixed at $K = 5$. Due to limited computational resources, we perform $15$ epochs for each fold, resulting in a total of $75$ epochs for all subsequent analyses. \\

\subsubsection{Evaluation Metrics}

\noindent To evaluate the performance of the model, three main metrics are employed: accuracy, the Receiver Operating Characteristic (ROC) curve, and the Area Under the Curve (AUC). Accuracy measures the overall percentage of correctly classified jet images across all classes, providing a general measure of predictive correctness. The ROC curve characterizes the trade-off between true positive rare (TPR) and false positive rate (FPR) across varying classification thresholds, offering insight into the model's ability to distinguish between classes. These metrics are widely used in high-energy physics for their ability to compare signal versus background separation capabilities \cite{bradshaw2023guide, rocReview2006}. Given a particular class treated as the positive class (e.g., quark jets), and the other as negative (e.g., gluon jets), we define:
\begin{equation}
\text{TPR} = \frac{\text{TP}}{\text{TP} + \text{FN}} = \frac{M_{ii}}{\sum_j M_{ij}}; \quad \text{FPR} = \frac{\text{FP}}{\text{FP} + \text{TN}} = \frac{\sum_{j \neq i} M_{ji}}{\sum_{k \neq i} \sum_{l} M_{kl}} ,
\end{equation}
where \footnote{TP$\equiv$ True Positive, FP$\equiv$ False Positive, TN$\equiv$ True Negative and FN$\equiv$ False Negative.} $M_{ii}$ denotes true positives for class $i$, and $M_{ji}$ corresponds to false positives-samples from class $j \neq i$ misclassified as class $i$. These values are computed for each class in a one-vs-rest manner. And then, as mentioned before, the ROC curve is constructed by plotting TPR against FPR as the classification threshold varies.. And finally, AUC is a single number\footnote{the closer it is to 1, the better the model is at distinguishing between classes.} that summarize the ROC curve, with higher values indicating better discrimination capability. Together, these metrics give a clear picture of model performance and its reliability in jet classification tasks. \\

\subsection{Component-wise analysis} \label{sec:S2.3}
\noindent In this work, a component-wise analysis is performed to assess the contribution of each model component to the performance of the EM. This is done by evaluating the classification results when each of the constituent networks within the EM is trained and tested independently, by disabling the concatenation between the \emph{ResNet50} and \emph{InceptionV3} outputs. \\

\noindent Note that even when the concatenation layer in the EM architecture is bypassed, the fusion layer still projects the output from either \emph{ResNet50} or \emph{InceptionV3} to a $512$ dimensional representation (figure \ref{fig:ensemble}). Therefore, this component-wise analysis can also be viewed as an ablation study, providing insights into how the individual parts of the network contribute to its overall functionality \cite{meyes2019ablation}. The goal is to determine how much discriminative power each individual architecture contributes, and how their combination improves jet classification. The corresponding results can be found in Section \ref{sec:S4.1} for binary classification tasks and in Section \ref{sec:S4.2} for the multi-class scenario.

\section{Results} \label{sec:S4}

\noindent This section presents the outcomes of applying the proposed EM to both binary (Section \ref{sec:S4.1}) and multi-class (Section \ref{sec:S4.2}) jet classification tasks using the JetNet dataset. Performance is evaluated in terms of classification accuracy and AUC metrics, as well as through ROC curve analysis. All metrics are reported under a 5-fold cross-validation protocol, with $15$ epochs of training for each fold. To better understand the contribution of each component within the ensemble, we include component-wise analysis that evaluates the individual performance of \textit{ResNet50} and \textit{InceptionV3}, within the EM architecture.\\

\subsection{Binary jet Classification} \label{sec:S4.1}

\noindent This classification task focuses on distinguishing between the QCD background (or the gluon-derived jets ($g$-jets)) and jets originating from other particle types (light quarks, top quarks, $W$ and $Z$ bosons). The models were trained and evaluated using 5-fold cross-validation on a balanced subset of the dataset, where each class contains an equal number of samples to avoid bias. The performances of each individual architecture, as well as the proposed EM for binary classification are summarized in tables~\ref{tab:gq-results}-\ref{tab:gz_results}, which report the average accuracies (training/validation and testing) and AUC obtained across the five cross-validation folds. The computations in this paper were run on the FASRC Cannon cluster, supported by the FAS Division of Science Research Computing Group at Harvard University.\\

\begin{table}[htpb]
\centering
\begin{tabular}{lccccc}
\hline 
Model & Training Acc. & Validation Acc. & Testing Acc. &  AUC & Time (min) \\ 
\hline
ResNet50 & 0.8090 & 0.7115 & 0.7854 & 0.858 & \textbf{667}\\
InceptionV3 & 0.8188 & 0.7242 & 0.7854 & 0.859 & 678\\
Ensemble &\textbf{ 0.8611} & \textbf{0.7632} & \textbf{0.7940} & \textbf{0.871} & 710 \\ 
\hline 
\end{tabular}
\caption{Performance metrics for binary classification models for $g$-jet vs. $q$-jet. }
\label{tab:gq-results}
\end{table}

\begin{table}[htpb]
\centering
\begin{tabular}{lccccc}
\hline 
Model & Training Acc. & Validation Acc. & Testing Acc. & AUC & Time (min) \\ 
\hline
ResNet50 & 0.9055 & 0.8332 & 0.8671 & 0.941 & \textbf{638}\\ 
InceptionV3 & 0.8991 & 0.8480 & 0.8678 & 0.942 & 671\\ 
Ensemble & \textbf{0.9164} & \textbf{0.8593} & \textbf{0.8744} & \textbf{0.950} & 687 \\ 
\hline 
\end{tabular}
\caption{Performance metrics for binary classification models for $g$-jet vs. $t$-jet.}
\label{tab:gt_results}
\end{table}

\begin{table}[htpb]
\centering
\begin{tabular}{lccccc}
\hline 
Model & Training Acc. & Validation Acc. & Testing Acc. & AUC & Time (min) \\ 
\hline
ResNet50 & 0.9265 & 0.8704 & 0.9102 & 0.968 & \textbf{587}\\ 
InceptionV3 & 0.9200 & 0.9037 & 0.9016 & 0.967 & 684\\ 
Ensemble & \textbf{0.9515} & \textbf{0.9083} & \textbf{0.9175} & \textbf{0.973} & 698 \\ 
\hline 
\end{tabular}
\caption{Performance metrics for binary classification models for $g$-jet vs. $W$-jet.}
\label{tab:gw_results}
\end{table}

\begin{table}[htpb]
\centering
\begin{tabular}{lccccc}
\hline 
Model & Training Acc. & Validation Acc. & Testing Acc. & AUC & Time (min) \\ 
\hline
ResNet50 & 0.9285 & 0.8896 & 0.9079 & 0.968 & \textbf{583}\\ 
InceptionV3 & 0.9270 & 0.8762 & 0.9016 & 0.965 & 597\\ 
Ensemble & \textbf{0.9378} & \textbf{0.9002} & \textbf{0.9122} & \textbf{0.974} & 632 \\ 
\hline 
\end{tabular}
\caption{Performance metrics for binary classification models for $g$-jet vs. $Z$-jet.}
\label{tab:gz_results}
\end{table}

\noindent Table~\ref{tab:gq-results} to table~\ref{tab:gz_results}, where \emph{Time} column refers to the time needed to train on the FASRC Cannon cluster, show that when \textit{InceptionV3} was removed from the ensemble, \textit{ResNet50} achieves average validation accuracies ranging from 0.71 to 0.88, testing accuracies from 0.78 to 0.91, and AUC score between 0.85 and 0.96. On the other hand, when \textit{ResNet50} was removed from the ensemble, \textit{InceptionV3} achieves validation accuracies ranging from 0.72 to 0.90, testing accuracies from 0.78 to 0.91, and AUC score between 0.85 and 0.96. The EM outperforms both individual models, achieving more stable and higher results: validation accuracy ranging from 0.76 to 0.90, testing accuracy from 0.79 to 0.91, and  AUC scores between 0.87 and 0.97. This highlights the performance gains from the ensemble approach. While both models exhibit strong classification capabilities, their results also show variability across folds, suggesting some sensitivity to data partitioning. In contrast, the EM consistently outperforms both and exhibits narrower performance fluctuations, indicating improved robustness and generalization.\\

\subsubsection{Binary ROC curves} \label{sec:S4.1.1}

\begin{figure}[h]
    \centering
    \begin{subfigure}{0.49\textwidth}
        \centering
        \includegraphics[width=\textwidth]{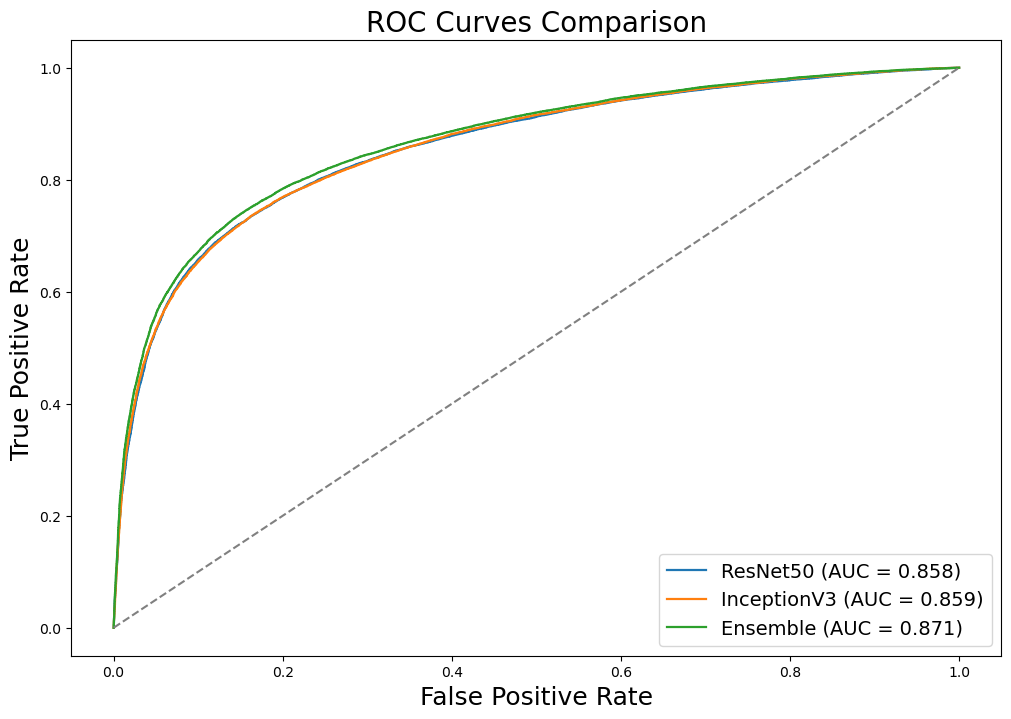} 
        \caption{ } 
        \label{fig:image1}
    \end{subfigure}
    \hfill
    \begin{subfigure}{0.49\textwidth}
        \centering
        \includegraphics[width=\textwidth]{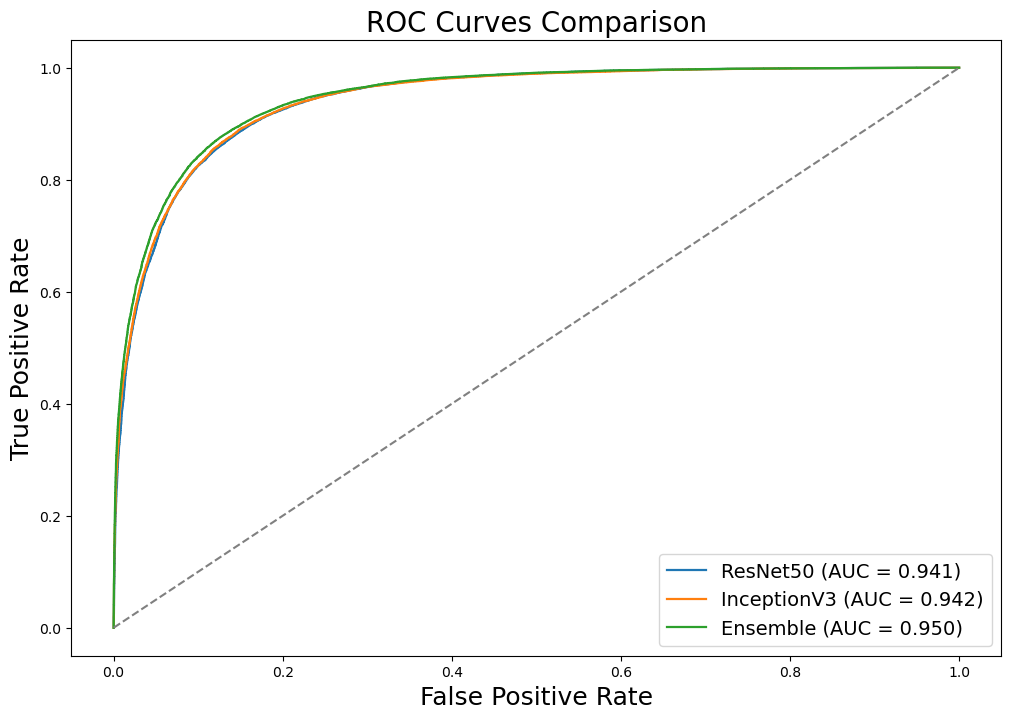} 
        \caption{}
        \label{fig:image2}
    \end{subfigure}
    \hfill
    \begin{subfigure}{0.49\textwidth}
        \centering
        \includegraphics[width=\textwidth]{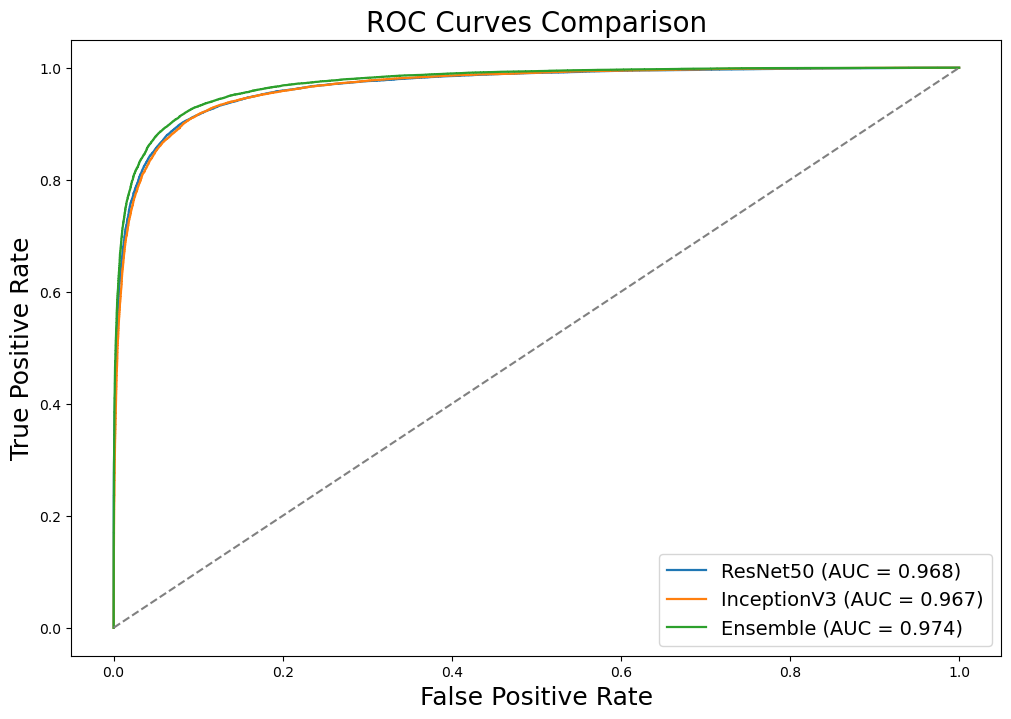} 
        \caption{}
        \label{fig:image3}
    \end{subfigure}
    \hfill
    \begin{subfigure}{0.49\textwidth}
        \centering
        \includegraphics[width=\textwidth]{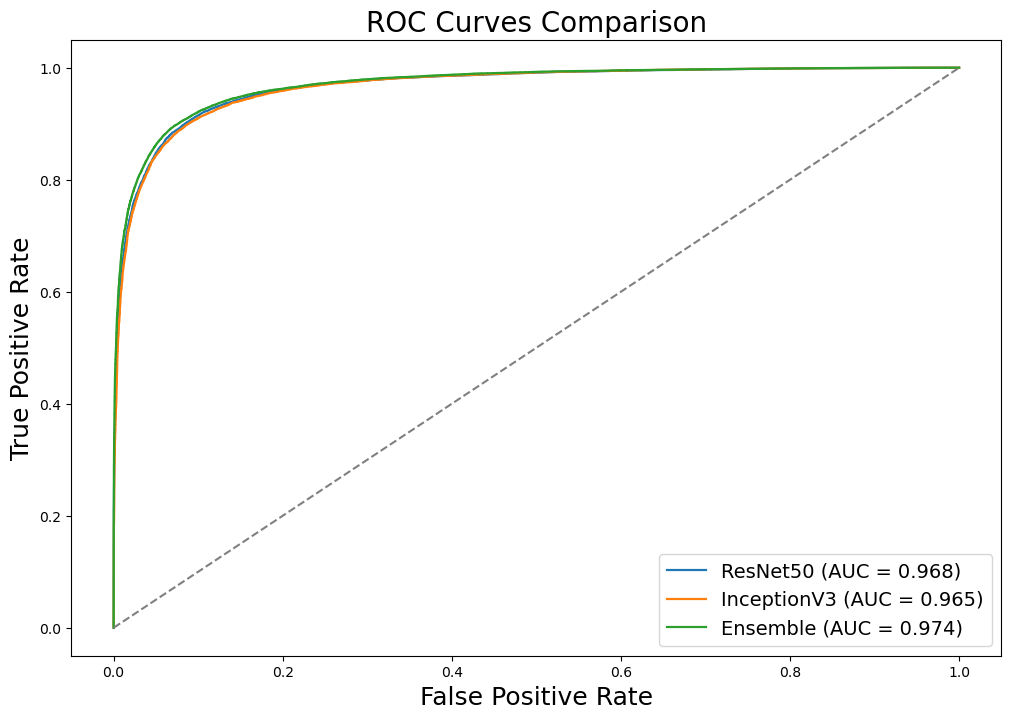} 
        \caption{}
        \label{fig:image4}
    \end{subfigure}
    \caption{ROC curve plots for binary classification tasks. Plot (a) correspond to the ROC curves for $g$-jet vs. $q$-jet case, plot (b) correspond to the ROC curves for $g$-jet vs. $t$-jet case, plot (c) correspond to the ROC curves for $g$-jet vs. $W$-jet case, and plot (d) correspond to the ROC curves for $g$-jet vs. $Z$-jet case. }
    \label{fig:binary_ROC-curves}
\end{figure}

\noindent To further illustrate the comparative performance of the individual models and the EM, figure~\ref{fig:binary_ROC-curves} presents the ROC curves obtained during the binary classification tasks. These plots provide a visual assessment of the ability of each model to distinguish between: gluon and quark jets (figure~\ref{fig:image1}), gluon and top-quark jets (figure~\ref{fig:image2}), gluon and $W$-boson jets (figure~\ref{fig:image3}), and gluon and $Z$-boson jets (figure~\ref{fig:image4}), across different decision thresholds.

\noindent As part of the component-wise study, the ROC curves highlight that the EM consistently achieves higher true positive rates (TPR) at lower false positive rates (FPR),\footnote{These thresholds are critical in HEP applications, where misclassifications can significantly impact signal–background separation.} reinforcing the performance gains observed in terms of accuracy and AUC.\\

\subsection{Multi-Class jet Classification} \label{sec:S4.2}

\noindent The objective is to correctly identify jets originating from the five distinct particles types, those that can be generated from the JetNet dataset. This task presents greater complexity compared to binary classification, as it requires the model to learn finer distinctions among classes with overlapping physical signatures. The models were also trained and evaluated using the same $5$-fold cross-validation on balanced dataset, ensuring equal representation of each jet class to prevent bias. As in binary case, we evaluate performance using accuracy, AUC and ROC curves. The results are presented in the table~\ref{tab:multiclass_results}, where one can observe and compare the performance of ResNet50 (when InceptionV3 was removed), InceptionV3 (when InceptionV3 was removed), and the proposed EM. \\

\begin{table}[h]
\centering
\begin{tabular}{lccccc}
\hline
Model & Training Acc. & Validation Acc. & Testing Acc. & Avg. AUC & Time (min) \\
\hline
ResNet50 & 0.7383 & 0.7149 & 0.7248 & 0.920 & \textbf{2009}\\ 
InceptionV3 & 0.7525 & 0.7217 & 0.7369 & 0.923 & 2013\\ 
Ensemble & \textbf{0.7868} & \textbf{0.7298} & \textbf{0.7508} & \textbf{0.935} & 2052\\ 
\hline
\end{tabular}
\caption{Performance metrics for multi-class classification with pretrained models.}
\label{tab:multiclass_results}
\end{table}

\noindent Testing accuracy across the models ranges from $0.72$ to $0.75$, while the average AUC scores from $0.92$ to $0.93$. These values reflect the increased complexity of the multi-class problem compared to the binary. \textit{ResNet50} achieves the lowest ranging values for testing accuracy and average AUC, $0.72$ and $0.92$ respectively. \textit{InceptionV3} performs slightly better in this setting, with a testing accuracy of $0.74$ and an AUC of $0.93$. On the other side, the EM achieves a testing accuracy of $0.75$ and an AUC of $0.93$, which confirm the advantage of the ensemble strategy. It consistently outperforms both base models, not only in terms of predictive accuracy but also in generalization stability across folds. This component-wise study also confirms that neither \textit{ResNet50} nor \textit{InceptionV3} alone is sufficient to fully capture the range of features necessary for robust multi-class classification. However, when combined, they enhance the discriminative power of the model. \\

\subsubsection{Multi-class ROC curves}\label{sec:S4.2.1}
\begin{figure}[H]
    \centering
    \begin{subfigure}{0.48\textwidth}
        \centering
        \includegraphics[width=\textwidth]{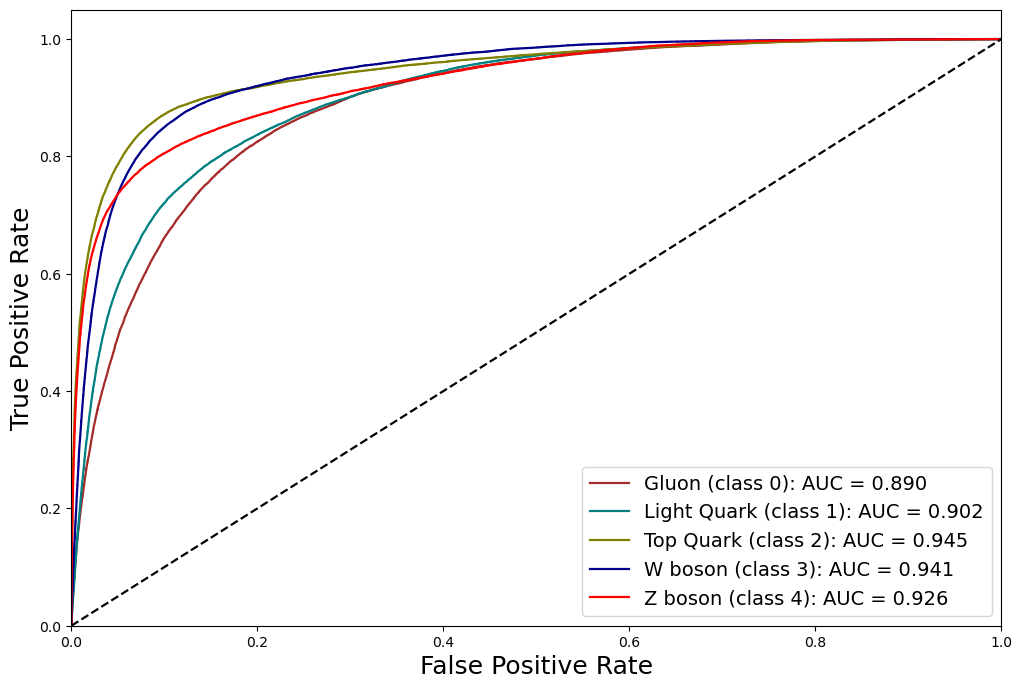} 
        \caption{ ResNet50 } 
        \label{fig:m-image1}
    \end{subfigure}
    \hfill
    \begin{subfigure}{0.48\textwidth}
        \centering
        \includegraphics[width=\textwidth]{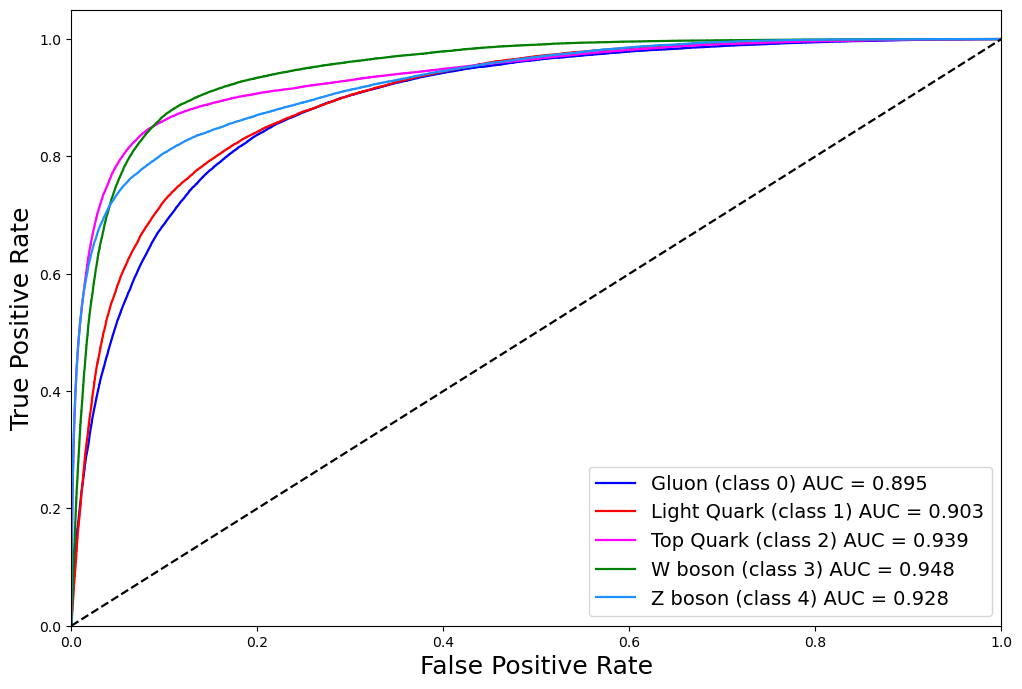} 
        \caption{InceptionV3}
        \label{fig:m-image2}
    \end{subfigure}
    \hfill
    \begin{subfigure}{0.60\textwidth}
        \centering
        \includegraphics[width=\textwidth]{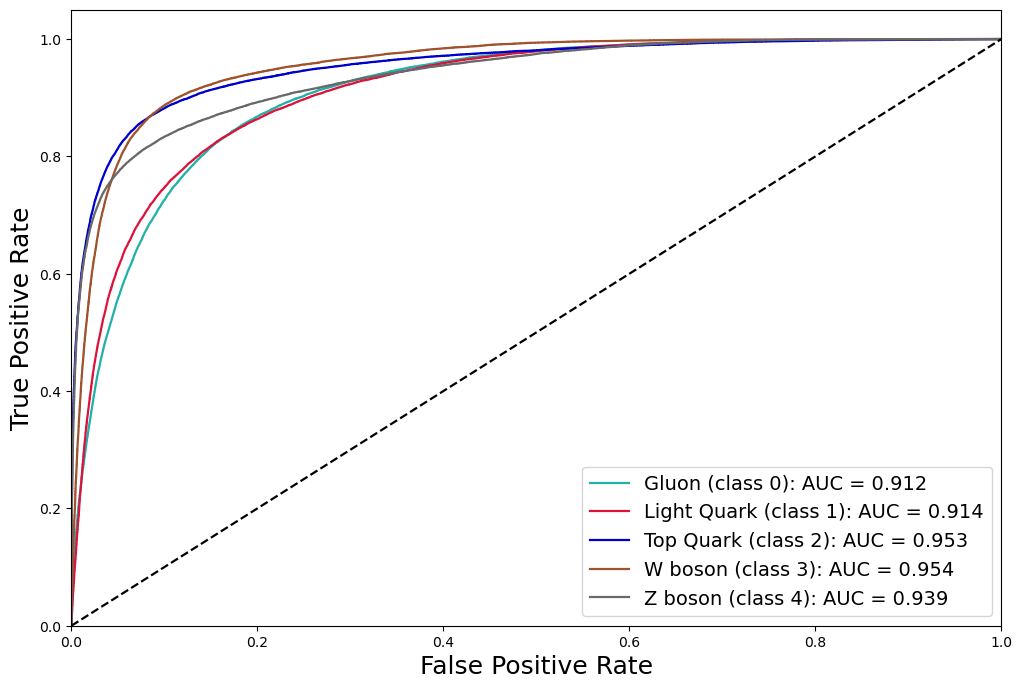} 
        \caption{Ensemble}
        \label{fig:m-image3}
    \end{subfigure}
    \caption{ROC curve plots for multi-class classification tasks. The plots display the one-vs-rest ROC performance for each jet class: Gluon, Light Quark, Top Quark, W Boson, and Z Boson. The background class in all one-vs-rest evaluations corresponds to gluon jets.}
    
    \label{fig:multi_ROC-curves}
\end{figure}

\noindent To complement the quantitative metrics reported before, figure~\ref{fig:multi_ROC-curves} displays the ROC curves corresponding to the multi-class scenario, where the performance of each model is evaluated independently for each jet category. They were generated for each class in a one-vs-rest approach and help to visualize how well each model discriminates among each class from the rest, which is particularly useful in cases where classes share similar jet substructures. 

\noindent The ROC curve plots presented in figure \ref{fig:multi_ROC-curves} provide a class-by-class visualization of the models performance for the five jet categories. From the individual standpoints, \textit{ResNet50} (figure \ref{fig:m-image1}) exhibits relatively sharper ROC profiles for classes with compact and localized radiation patterns, such as $q$-jet (class 1). This aligns with its architectural design, which favors depth and localized feature extraction through residual learning. However, its curves for more complex jet structures\footnote{specifically $W$ and $Z$ jets (class 3 and class 4).} show slightly diminished separability, likely due to the limited multi-scale processing capacity of the network. \textit{InceptionV3} (figure \ref{fig:m-image2}), in contrast, produces more balanced ROC curves across all classes, particularly improving separability for classes such as $t$-jet (class 2) and bosonic jets (classes 3 and 4), which exhibit more distributed substructure. This is attributable to its architectural use of parallel convolutions with varying receptive fields, which better capture multi-scale energy deposits\footnote{characteristic of boosted decays.}. 
From the EM (figure \ref{fig:m-image3}), its ROC curves remain closer to the top-left corner in the region of low false-positive rate, indicating a stronger ability to detect true positives early while keeping misclassifications low. In particular, t-jets and bosonic jets achieve the most pronounced curves, suggesting that the EM effectively captures their characteristic profiles and substructures. For q-jets curve, it also preserve the sharp discrimination seen in figures \ref{fig:m-image1} and \ref{fig:m-image2}, while for the g-jets Roc curve a significant improvement can be seen beyond either individual model due to the combination of residual depth for local pattern recognition and inception-based feature aggregation. Overall, the results in figure \ref{fig:m-image3} illustrate how merging complementary architectural strengths allows the model to maintain high discriminative power across jet categories with different structural features.

\subsubsection{Feature analysis}\label{sec:4.2.2}
\paragraph{Stability of Results:} In evaluating the performance gain of the EM over constituent architectures, it is crucial to address the statistical stability of the observed metrics. One might reasonably argue whether the apparent improvements in accuracy and AUC could arise from random fluctuations due to training with different initializations. To investigate this, repeated training runs for multi-class classification were performed using different random seed values, with the same $5$-fold cross-validation and $15$ epochs of training for each fold. The results are as follows:
\begin{itemize}
\item \textbf{\emph{ResNet50}:} Mean AUC = $0.9257$, Standard deviation = $0.0004$
\item \textbf{\emph{InceptionV3}:} Mean AUC = $0.9298$, Standard deviation = $0.0002$
\item \textbf{Ensemble Model:} Mean AUC = $0.9342$, Standard deviation = $0.0002$
\end{itemize}
These small standard deviations indicate that the models yield highly stable performance across different random initializations. Furthermore, to quantify the statistical significance of the performance differences, we performed unpaired two-sample $t$-tests between the model accuracies:
\begin{itemize}
\item \textbf{\emph{ResNet50} vs Ensemble:} $t=-35.6655$, $p=3.57\times 10^{-8}$
\item \textbf{\emph{InceptionV3} vs Ensemble:} $t=-28.8874$, $p=2.2\times 10^{-9}$
\item \textbf{\emph{ResNet50} vs \emph{InceptionV3}:} $t=-17.0721$, $p=2.76\times 10^{-6}$
\end{itemize}
The extremely low $p$-values in all comparisons provide strong statistical evidence that the differences in mean performance across models are not due to chance. In particular, the ensemble model's superiority is statistically significant when compared to both \emph{ResNet50} and \emph{InceptionV3} individually. This reinforces our conclusion that the ensemble architecture offers a consistent and reliable performance edge, not merely a product of stochastic fluctuations during training.
\begin{table}[h]
	\centering
	\begin{tabular}{lccccc}
		\hline
		Model & Training Acc. & Validation Acc. & Testing Acc. & Avg. AUC & Time (min) \\
		\hline
		ResNet50 & 0.6892 & 0.6165 & 0.6243 & 0.867 & \textbf{1843}\\ 
		InceptionV3 & 0.7003 & 0.6401 & 0.6588 & 0.872 & 1905\\ 
		Ensemble & \textbf{0.7379} & \textbf{0.6801} & \textbf{0.7007} & \textbf{0.902} & 1984\\ 
		\hline
	\end{tabular}
	\caption{Performance metrics for multi-class classification with No-pretrained models.}
	\label{tab:multiclass_results-NP}
\end{table}
\paragraph{Pre-training advantage:} We initialized both \emph{ResNet50} and \emph{InceptionV3} with weights pre-trained on the ImageNet dataset \cite{deng2009imagenet}, based on our empirical observation that pre-training consistently leads to improved performance on the jet classification task when trained for the same number of epochs. As shown in table \ref{tab:multiclass_results-NP}, models initialized with random weights only partially reach the performance of their pre-trained counterparts under identical training schedules. This highlights the fact that while the architecture, particularly the presence of convolutional layers, is crucial for jet-image tagging, initializing with pre-trained weights offers a practical advantage by significantly reducing training time and computational cost.

\begin{figure} [tbph]
    \centering
    \begin{subfigure}{0.9\textwidth}
        \centering
        \includegraphics[width=0.95\textwidth]{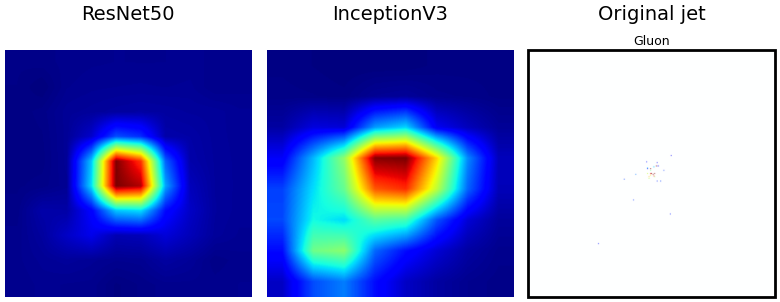} 
    \end{subfigure}
    \begin{subfigure}{0.9\textwidth}
        \centering
        \includegraphics[width=0.95\textwidth]{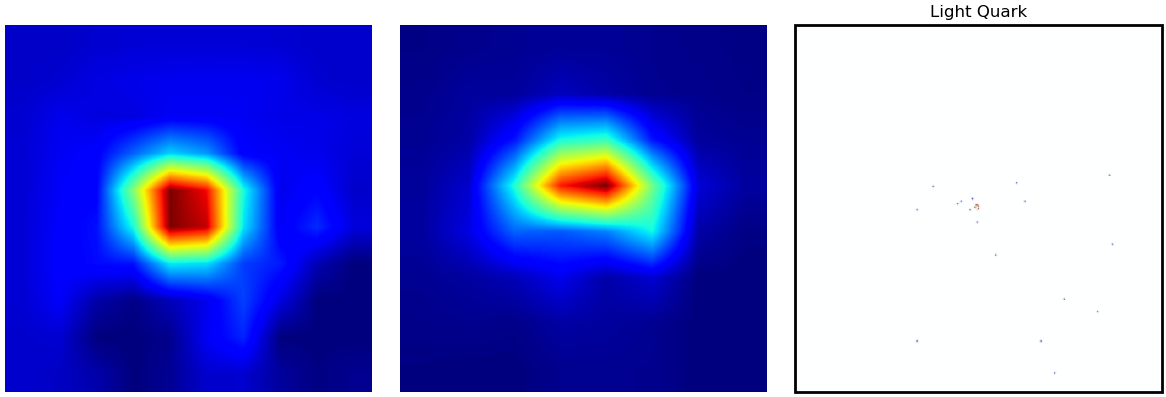} 
    \end{subfigure}
    \begin{subfigure}{0.9\textwidth}
        \centering
        \includegraphics[width=0.95\textwidth]{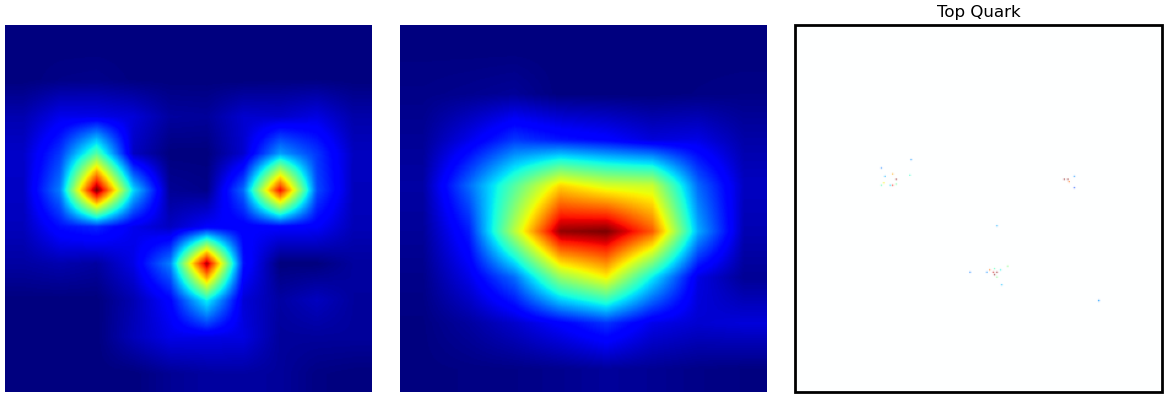} 
    \end{subfigure}
    
    \begin{subfigure}{0.9\textwidth}
        \centering
        \includegraphics[width=0.95\textwidth]{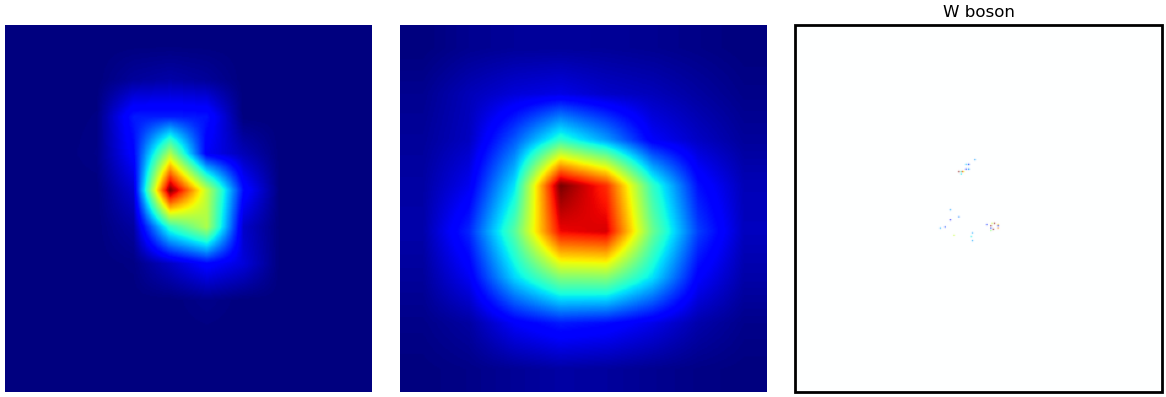}
    \end{subfigure}
    \hfill
    \begin{subfigure}{0.9\textwidth}
        \centering
        \includegraphics[width=0.95\textwidth]{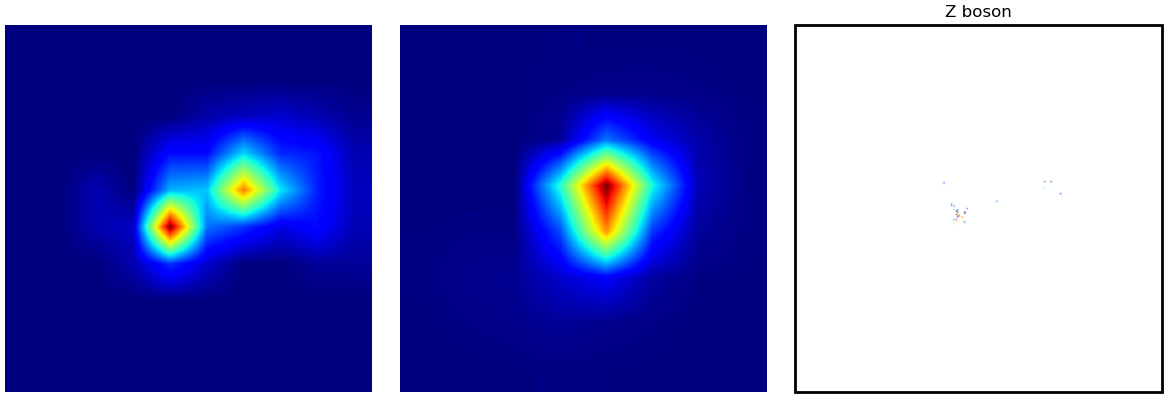} 
    \end{subfigure}
    \hfill
    
    \caption{Grad-CAM Analysis of \textit{ResNet50} vs \textit{InceptionV3}.}
    \label{fig:Heat-map_jet_image}
\end{figure}

\paragraph{Feature adaption:} \textit{ResNet50’}s deep residual structure is known to facilitate the preservation and refinement of localized spatial features across many layers, while InceptionV3’s multi-branch architecture is explicitly designed to capture spatial information through its use of parallel convolutional layers. To substantiate this for the present scenario, we use the Gradient-weighted Class Activation Mapping (Grad-CAM) \cite{Selvaraju_2019}, which uses the gradients of any target concept flowing into the final convolutional layer to produce a heat-map highlighting the important regions in the image for predicting the concept. This method effectively assigns importance scores to spatial locations by performing a weighted combination of the forward activation maps, where the weights are the average gradients associated with the target class. The resulting heatmap is color-coded such that warmer colors (e.g., red/yellow) indicate regions of high relevance, while cooler colors (e.g., blue) indicate low or no contribution to the model’s prediction. This Grad-CAM visualisation is performed during the component-wise analysis, so it required no extra steps in training of the networks. The resulting heatmaps, shown in figure \ref{fig:Heat-map_jet_image},  broadly confirms that \textit{ResNet50} is particularly effective at capturing localized jet features, as evidenced by concentrated activations around local clusters. In contrast, \textit{InceptionV3} captures broader spatial patterns across the jet image. The EM architecture benefits from combining these complementary strengths through feature-level fusion, allowing it to construct a more discriminative internal representation of jet images. This becomes particularly advantageous in binary classification tasks such as gluon vs. non-gluon tagging, where subtle differences in energy distribution can be challenging to resolve.

\section{Discussion and Conclusion} \label{sec:S5}

\noindent The results presented in this work suggest that ensemble deep learning models, when built using architecturally diverse CNNs, can offer substantial performance advantages in jet tagging while remaining computationally viable. Although the model requires higher computational cost than individual models, the trade-off is justified by the performance gains.

\noindent This observation can be further substantiated by the ROC curves presented in Sections \ref{sec:S4.1.1} and \ref{sec:S4.2.1} along with the t-test results and Grad-CAM analysis in Section \ref{sec:4.2.2},  which provide a class-by-class visualization of models discriminative performance. In the binary classification task, distinguishing $g$-jets from jets initiated by other particles, the EM achieved the highest performance, with testing accuracies of up to $0.91$ and AUC of $0.97$ for both the $g$-jet vs. $W$-jet as well as the $g$-jet vs. $t$-jet cases. This exceeds the performance of the standalone networks and indicating that combining localized feature extraction (from \textit{ResNet50}) and multi-scale spatial pattern recognition (from \textit{InceptionV3}) leads to more discriminative and generalizable representations. ROC curves from the component-wise study (plots (a) to (d) in figure \ref{fig:binary_ROC-curves}) in the binary case confirmed the advantage of the EM in maintaining higher TPR at lower FPR. Within the context of the ablation study for multiclass classification task, the ROC curves clearly illustrate the added value of the EM approach: while \textit{ResNet50} (figure~\ref{fig:m-image1}) demonstrates strong separability for classes with compact, localized energy deposits (e.g., quarks), and \textit{InceptionV3} (figure~\ref{fig:m-image2}) performs better on spatially extended classes (e.g., $W$ and $Z$ bosons), the EM (figure~\ref{fig:m-image3}) consistently improves or maintains higher TPR across all classes. ROC analysis revealed that the EM offers better class separation across the board, achieving more reliable and generalized class boundaries even in overlapping regions of feature space. This reinforces the value of CNNs symmetry learning ability in handling the complexity of real jet substructures in HEP data. \\

\noindent Importantly, these performance gains were achieved with only a moderate increase in computational cost. While training time (represented as Time in the last column of tables \ref{tab:gq-results} to \ref{tab:multiclass_results}) for the ensemble was higher due to the dual-model architecture, this overhead is acceptable for offline analysis tasks and potentially adaptable for real-time system with further optimization. The training process for the EM completed in approximately $2052$ minutes for all $5$-classes of jet data, which is reasonable considering the size of the dataset and the cross-validation protocol implemented.

\noindent The balance achieved between accuracy, AUC performance, and inference time supports EM's potential for integration into physics analyses, such as those conducted in LHC experiments and searches for physics beyond the Standard Model. We are currently working on extending the Ensemble Model to incorporate a larger set of jet features, making it more flexible and better suited for information-rich datasets such as JetClass. Looking forward, future research could incorporate additional model diversity, including but not limited to transformer-based vision architectures, with the aim of optimizing computational efficiency and extending the applicability of the model to real-time jet tagging scenarios in experimental setups. The framework could also be extended to semi-supervised learning paradigms, enhancing its relevance for data-driven discoveries in modern collider experiments. As an immediate next step, we plan to incorporate multiple histograms representing various jet-level observables as composite inputs to the EM module, offering a richer representation of jet image geometry.

\section*{Acknowledgment}

\noindent This work was supported by US NSF Award 2334265. The authors thank the artificial intelligence imaging group (aiig) laboratory at the University of Puerto Rico, Mayaguez for the computational facility and the investigators of the NSF Institute for Artificial Intelligence and Fundamental Interactions for their valuable insights during the course of this research. The authors also acknowledge the FASRC Cannon cluster supported by the FAS Division of Science Research Computing Group at Harvard University.

\appendix
\section{Quick R\'esum\'e: \emph{ResNet50} \& \emph{InceptionV3}} \label{append:S1}

\noindent To effectively capture the multi-scale and hierarchical substructures present in jet images, we integrate two special types of convolutional neural network (CNN) architectures with distinct yet complementary design principles. For EM, we use pretrained weights for both of these models on the ImageNet dataset \cite{deng2009imagenet}. A general overview of both architectures follows below.
\begin{figure}[H]
\includegraphics[width=12cm]{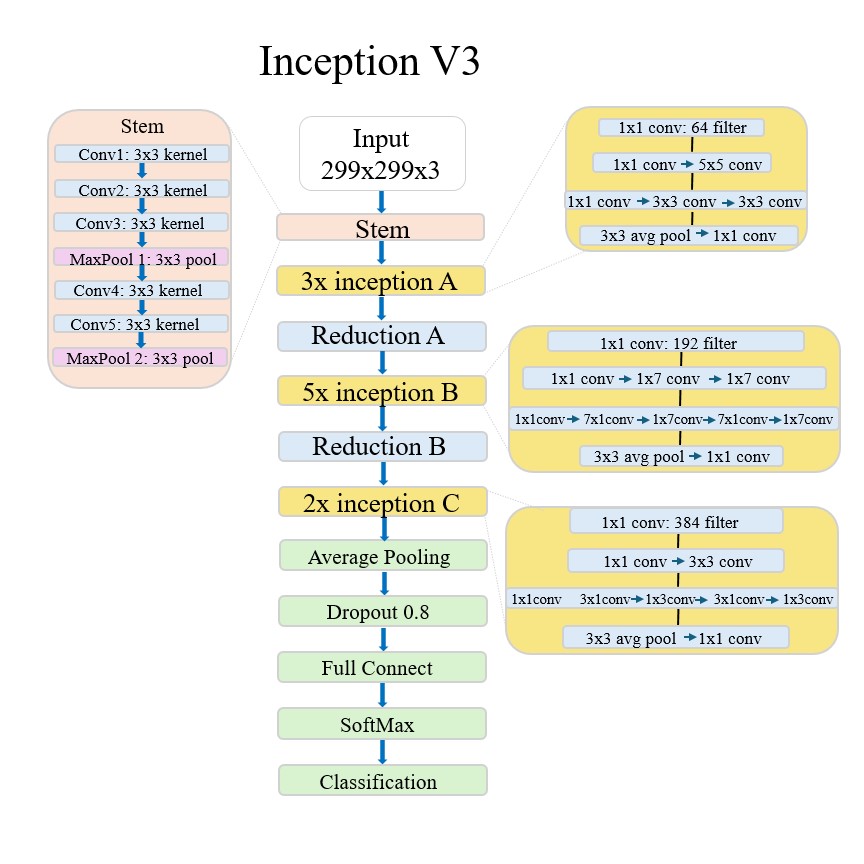}
\centering
\caption{Schematic overview of the \textit{InceptionV3} architecture used in this study. The model processes $299\times299\times3$ input images through a stem block, followed by multiple inception modules (A,B,C) and reduction blocks. Each inception module combines convolutions of different kernel sizes to capture multi-scale spatial features.}
\label{fig:Inception}
\end{figure}
\paragraph{InceptionV3} This is a 48-layer network developed by Szegedy et al. It employs multiple filter sizes and asymmetric convolutions, making it well-suited for recognizing diverse spatial patterns and capturing multi-scale features within jet images \cite{szegedy2016rethinking}. The InceptionV3, shown in figure \ref{fig:Inception}, is a CNN optimized for multi-scale feature extraction with a larger input size of $299 \times 299 \times 3$. The initial $Stem$ block consists of multiple $3 \times 3$ convolutions and max pooling layers, acting as a feature extractor before Inception modules. The main body of the network consists of three types of Inception modules (yellow blocks in the Inception diagram), with factorized $7\times7$ and $5\times5$ convolutions into asymmetric $1\times7$ + $7\times1$ operations, interleaved with reduction blocks. The final stage includes a 2048-D feature vector post average pooling, a dropout layer (rate = 0.8) to mitigate overfitting, a fully connected layer, followed by softmax activation for probabilistic classification. Its ability to recognize spatially distributed energy deposits is critical for discriminating any-other-class/gluon jets \cite{cogan2015jet}.
\begin{figure}[H]
\includegraphics[width=7.5cm]{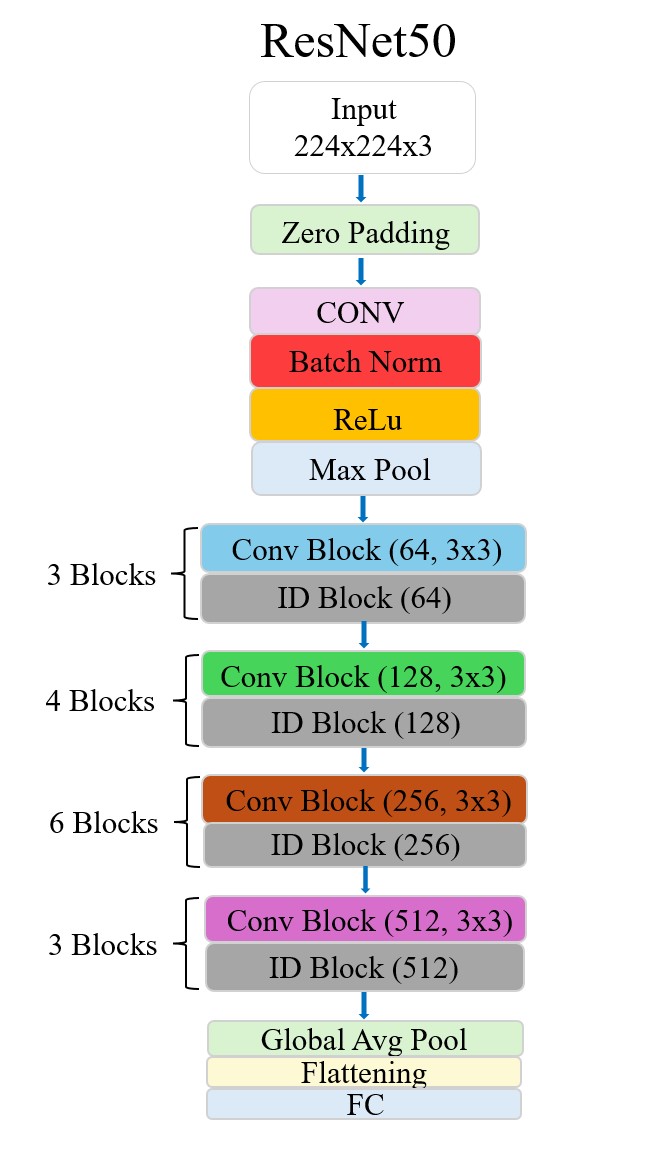}
\centering
\caption{Architecture of the \textit{ResNet50} model employed in this study. The network receives $224\times224\times3$ input images and passes them through an initial convolution and max pooling layer, followed by four stages of residual blocks composed of convolutional and identity mappings.}
\label{fig:resnet}
\end{figure}
\paragraph{ResNet50} This is a 50-layer deep residual network developed by \textit{He et al.} in 2016 \cite{he2016deep}, which introduced skip connections to mitigate the vanishing gradient problem in deep neural networks through residual learning \cite{baldi2015deep}, stabilizing training.
As shown in figure \ref{fig:resnet}, the \textit{ResNet50} architecture has an input size of $224 \times 224 \times 3$ (for both architectures, the number 3 in the input size represents the number of channels: RGB), followed by a zero-padding layer, and an initial stack of convolution, batch normalization, ReLU activation, and Max Pooling layers. The main body of the network consists of four sequential stages of 49 convolutional layers with $3\times3$ kernels, organized into 16 residual blocks with identity mappings, concluding with a global average pooling (GAP) layer producing a 1024-D feature vector, flattening, and a fully connected (FC) layer for classification.

\bibliographystyle{unsrt}  
\bibliography{biblio} 
\end{document}